\documentclass[a4paper]{jpconf}
\usepackage{graphicx}
\usepackage[square,sort&compress]{natbib}
\begin{document}
\title{The fainter the better: cataclysmic variable stars from the Sloan Digital Sky Survey}
\author{John Southworth, B\,T\,G\"ansicke and T\,R\,Marsh}
\address{University of Warwick, Coventry CV4~7AL, UK}
\ead{J.K.Taylor@warwick.ac.uk}

\begin{abstract}
The Sloan Digital Sky Survey has identified a total of 212 cataclysmic variables, most of which are fainter than 18th magnitude. This is the deepest and most populous homogeneous sample of cataclysmic variables to date, and we are undertaking a project to characterise this population. We have found that the SDSS sample is dominated by a great ``silent majority'' of old and faint CVs. We detect, for the first time, a population spike at the minimum period of 80 min which has been predicted by theoretical studies for over a decade.
\end{abstract}

\section{Introduction}

Cataclysmic variables (CVs) are interacting binary stars containing a white dwarf accreting material from a low-mass main sequence star via an accretion disc or stream.  Theoretical studies have consistently predicted that the population of CVs  should be dominated by short-period systems, with many piling up up at a minimum period of 60--70 minutes (Kolb 1993; Politano 1996). Unfortunately, the known population of CVs, visciously biased by observational selection effects, bears no resemblance to theoretical predictions (Ritter \& Kolb 2003). We are therefore undertaking a project to characterise the population of CVs discovered by the Sloan Digital Sky Survey (SDSS) (Szkody et al., 2002-6). This sample was identified spectroscopically and extends to faint magnitudes, so is much less biased than all previous samples.

\begin{figure}[h]
\includegraphics[width=26pc]{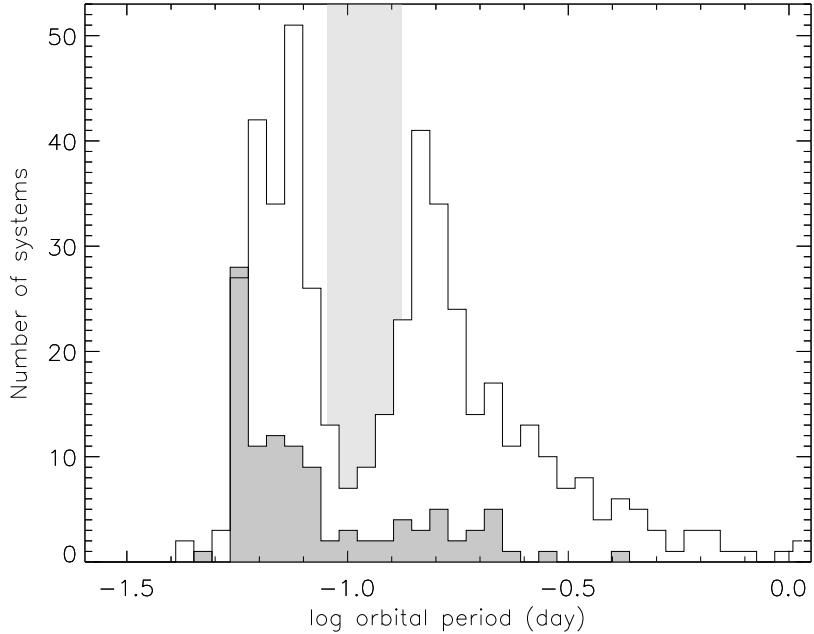}\hspace{2pc}%
\begin{minipage}[b]{10pc}
\caption{\label{fig:pd1}Orbital period distribution of the SDSS CVs
(grey histogram) compared to all other known CVs (white histogram).
The latter population has been taken from the catalogue of Ritter \&
Kolb (2003). The 2--3\,hour period gap in the known CV population is
shown with light shading. \\ \mbox{\ } \\ \mbox{\ } \\}
\end{minipage}
\end{figure}

We have identified the first CVs known to have a secondary star of brown dwarf mass (Southworth et al., 2006; Littlefair et al., 2006a, 2008), and for the first time are finding evidence for the long-predicted pile-up at the minimum period (G\"ansicke et al., 2008). The orbital period distribution of the SDSS CVs is compared to the distribution for the previously known CVs in Fig.\,\ref{fig:pd1}. Results for individual systems, based on observations with the VLT, NTT, WHT, INT and NOT, can be found in G\"ansicke et al.\ (2006, 2008), Southworth et al.\ (2006, 2007ab, 2008ab), Dillon et al.\ (2008) and Littlefair et al.\ (2006ab, 2007, 2008).

Our work on the SDSS population of CVs is intended to shed light on how these weird objects evolve. For contrast, we are also running a program to measure the orbital periods of a sample of pre-CVs which has been spectroscopically identified by the SDSS (Rebassa-Mansergas et al.\ 2007, 2008; Schreiber et al.\ 2008; see also G\"ansicke \& Schreiber 2003). Below we pick out two recent highlights of our characterisation of the SDSS CVs.

\section{SDSS J220553.98+115553.7: the pulsator which stopped}

SDSS\,J2205 has twice been observed to vary in brightness (Woudt \& Warner 2004; Szkody et al., 2007), with periods of 575\,s, 475\,s and 330\,s and amplitudes of $\sim$10\,mmag, which is typical of ZZ\,Ceti-type nonradial pulsations. Over two nights in 2008 August we obtained a light curve with the NTT. We were unable to detect these pulsations to a limit of 5\,mmag, but instead found a previously unseen photometric period of 44.8\,min. Our VLT spectroscopy yielded an orbital period of $82.83 \pm 0.09$\,min (Southworth et al., 2008a). The vanishing pulsation periods cannot be attributed to destructive interference or changes in accretion rate, but may be explicable by changes in the white dwarf temperature or the surface visibility of the pulsation modes.

\section{Triple-peaked Halpha emission from SDSS J003941.06$+$005427.5}

We obtained 29 spectra of SDSS\,J0039 over two nights in 2007 August using VLT/FORS2, finding an orbital period of $91.395 \pm 0.093$\,min. The spectra of this system show a remarkable and unique triple emission peak at H$\alpha$ (Southworth et al., 2008c). We have used the technique of Doppler tomography (Marsh \& Horne 1988) to generate velocity maps of the emission. These show that the inner peak moves in velocity with an amplitude of 180\,km\,s$^{-1}$, so cannot easily be attributed to either the white dwarf, secondary star, or accretion disc. Its existence remains a mystery. The H$\alpha$ Doppler maps also show that the accretion disc is very elliptical, which is not expected for a system in a state of very low mass transfer. Finally, there is also strong emission from the accretion disc in the [$+V_X$,$-V_Y$] quadrant, which has been seen in some ultra-compact binaries but is very odd for standard hydrogen-rich CVs. These aspects of SDSS\,J0039 defy explanation in the current picture of the properties of CVs, and demand further follow-up observations.

Doppler maps of the H$\alpha$ and He\,I 6678\,\AA\ emission lines from SDSS\,J0039 are shown in Figs.\ \ref{fig2} and \ref{fig3}. For comparison, we show Doppler maps from SDSS\,J2205 in Figs.\ \ref{f4} and \ref{f5}, which have a behaviour which is much more like that expected from short-period CVs: a circular accretion disc in H$\alpha$ emission and a bright spot visible in He\,I 6678\,\AA\ light where the mass transfer stream from the secondary star collides with the accretion disc. Overlaid on each Doppler map is a solid line indicating the Roche lobe of the secondary star in velocity space, crosses indicating the velocities of the centres of mass of the two stars and of the system, and dotted lines showing the velocity of the mass transfer stream as it follows a ballistic trajectory from the inner Lagrangian point.

\begin{figure}[h]
\begin{minipage}{18pc}
\includegraphics[width=17pc]{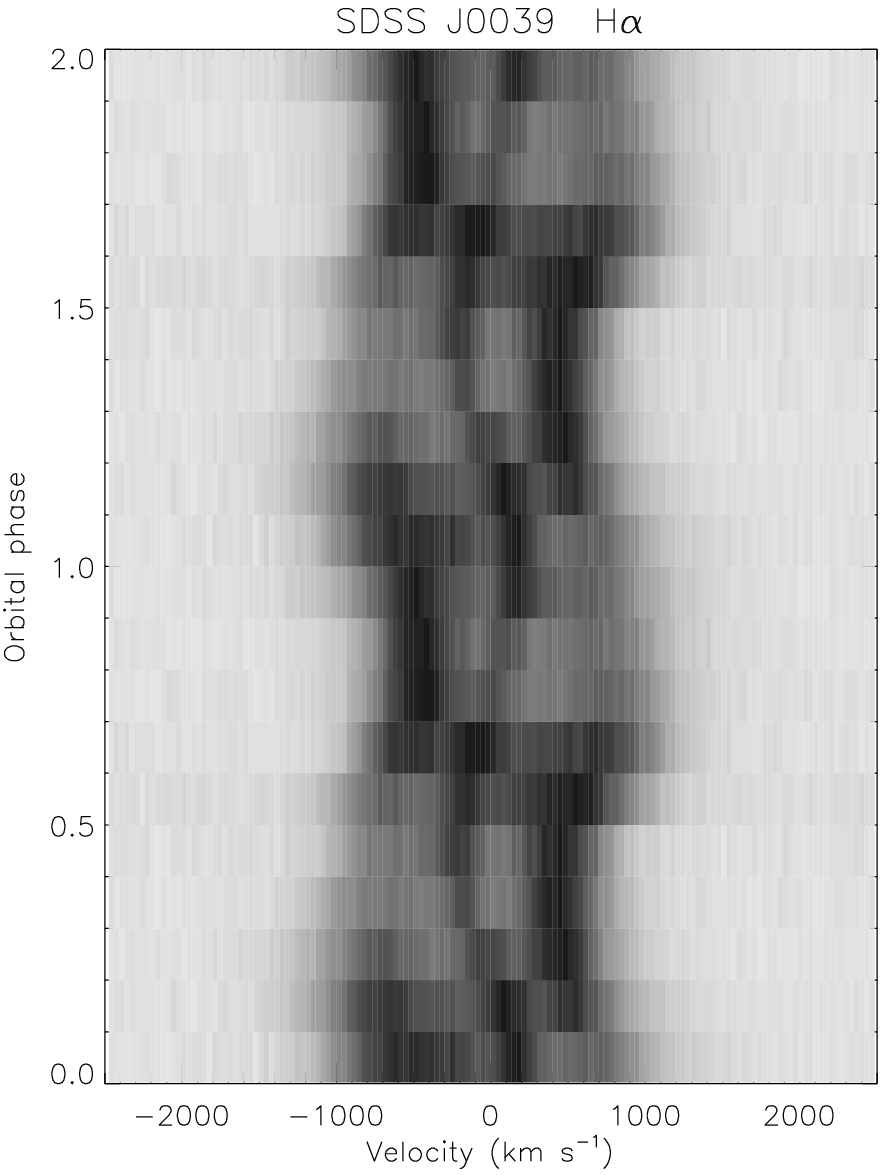}
\caption{\label{fig:0039ha} Trailed spectra of the H$\alpha$ emission line from
                            SDSS\,J0039. Two orbital phases are plotted for clarity.}
\end{minipage}\hspace{2pc}%
\begin{minipage}{18pc}
\includegraphics[width=17pc]{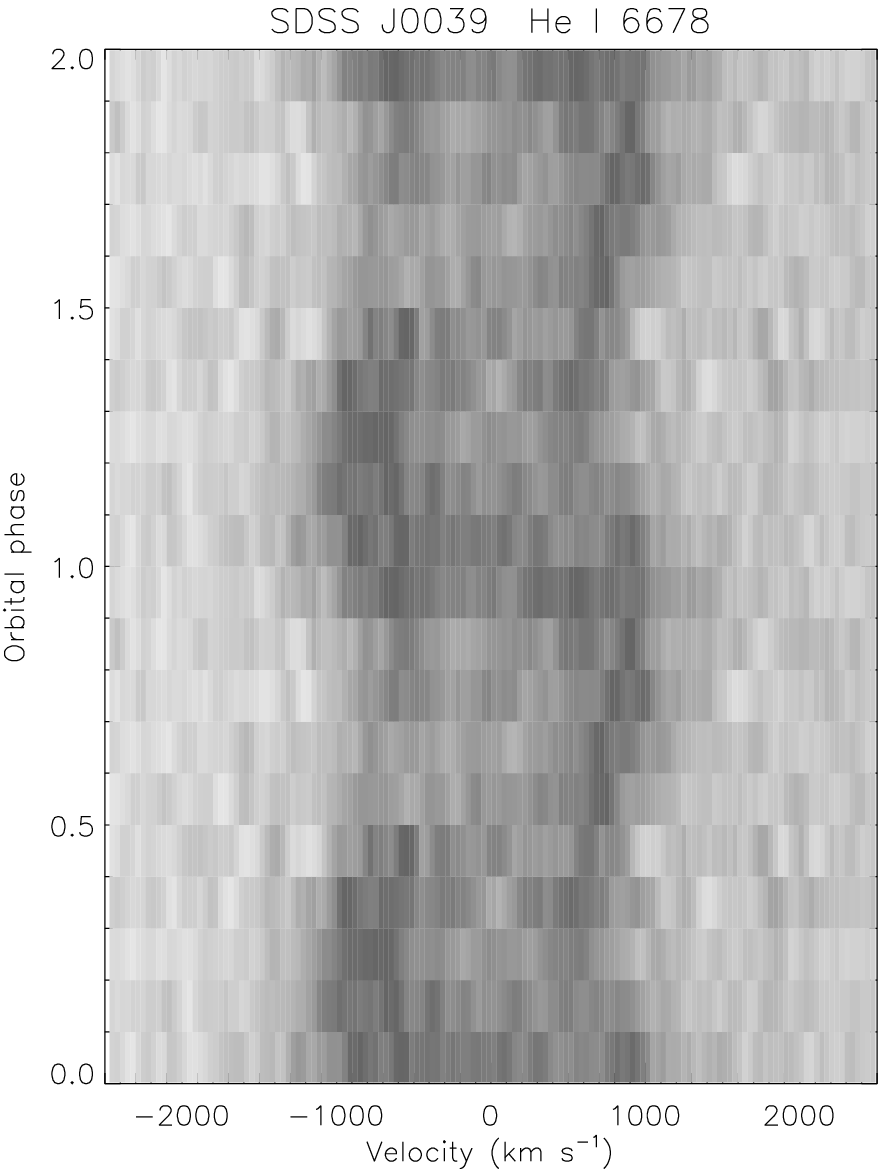}
\caption{\label{fig:0039he} Trailed spectra of the He\,I 6678\,\AA\ emission line from
                            SDSS\,J0039. The spectra have been smoothed for display.}
\end{minipage}
\end{figure}

\section*{References}

\begin{thereferences}

\item Dillon M, G\"ansicke B T, Aungwerojwit A, Rodr{\'{\i}}guez-Gil P, Marsh T R, Barros S C C, Szkody P, Brady S, Krajci T and Oksanen A 2008 {\it MNRAS} {\bf 386} 1568--1576

\item G\"ansicke B T et al.\ 2006 {\it MNRAS} {\bf 365} 969--976

\item G\"ansicke B T et al.\ 2008 {\it MNRAS} in preparation

\item Kolb U 1993 {\it A\&A} 271 149--166

\item Littlefair S, Dhillon V S, Marsh T R, G\"ansicke B T, Southworth J and Watson C A
      2006a {\it Science} {\bf 314} 1578--1580

\item Littlefair S, Dhillon V S, Marsh T R and G\"ansicke B T
      2006b {\it MNRAS} {\bf 371} 1435--1440

\item Littlefair S, Dhillon V S, Marsh T R, G\"ansicke B T, Baraffe I and Watson C A
      2007 {\it MNRAS} {\bf 381} 827--834

\item Littlefair S, Dhillon V S, Marsh T R, G\"ansicke B T, Southworth J, Baraffe I, Watson C A and Copperwheat C
      2008 {\it MNRAS} {\bf 388} 1582--1594

\item Marsh T R and Horne K D 1988 {\it MNRAS} {\bf 235} 269--286

\item Politano M, 1996 {\it ApJ} {\bf 465} 338--358

\item Rebassa-Mansergas A, G\"ansicke B T, Rodr{\'{\i}}guez-Gil P, Schreiber M R and Koester D
      2007 {\it MNRAS} {\bf 382} 1377--1393

\item Rebassa-Mansergas A et al.\ 2008, {\it Preprint} arXiv:0808.2148

\item Ritter H and Kolb U 2003 {\it A\&A} {\bf 404} 301--303

\item Schreiber M R and G\"ansicke B T 2003 {\it A\&A} {\bf 406} 305--321

\item Schreiber M R, G\"ansicke B T, Southworth J, Schwope A D and Koester D
      2008 {\it A\&A} {\bf 484} 441--450

\item Southworth J, G\"ansicke B T, Marsh T R, de Martino D, Hakala P, Littlefair S and Rodr{\'{\i}}guez-Gil P
      2006 {\it MNRAS} {\bf 373} 687--699

\item Southworth J, G\"ansicke B T, Marsh T R, de Martino D and Aungwerojwit A
      2007a {\it MNRAS} {\bf 378} 635--640

\item Southworth J, Marsh T R, G\"ansicke B T, Aungwerojwit A, Hakala P, de Martino D and Lehto H
      2007b {\it MNRAS} {\bf 382} 1145--1157

\item Southworth J, Townsley D M and G\"ansicke B T
      2008a {\it MNRAS} {\bf 388} 709--715

\item Southworth J, G\"ansicke B T, Marsh T R, Torres M A P, Steeghs D, Hakala P, Copperwheat C, Aungwerojwit A and Mukadam A
      2008b {\it Preprint} arXiv:0809.1753

\item Southworth J, Marsh T R and G\"ansicke B T 2008c in preparation

\item Szkody P, et al. 2002 {\it AJ} {\bf 123} 430--442

\item Szkody P, et al. 2003 {\it AJ} {\bf 126} 1499--1514

\item Szkody P, et al. 2004 {\it AJ} {\bf 128} 1882--1893

\item Szkody P, et al. 2005 {\it AJ} {\bf 129} 2386--2399

\item Szkody P, et al. 2006 {\it AJ} {\bf 131} 973--983

\item Szkody P, et al. 2007 {\it ApJ} {\bf 658} 1188--1195

\item Woudt P, Warner B 2002 {\it Ap\&SS} {\bf 282} 433--438

\end{thereferences}

\begin{figure}[t]
\begin{minipage}{18pc}
\includegraphics[width=18pc]{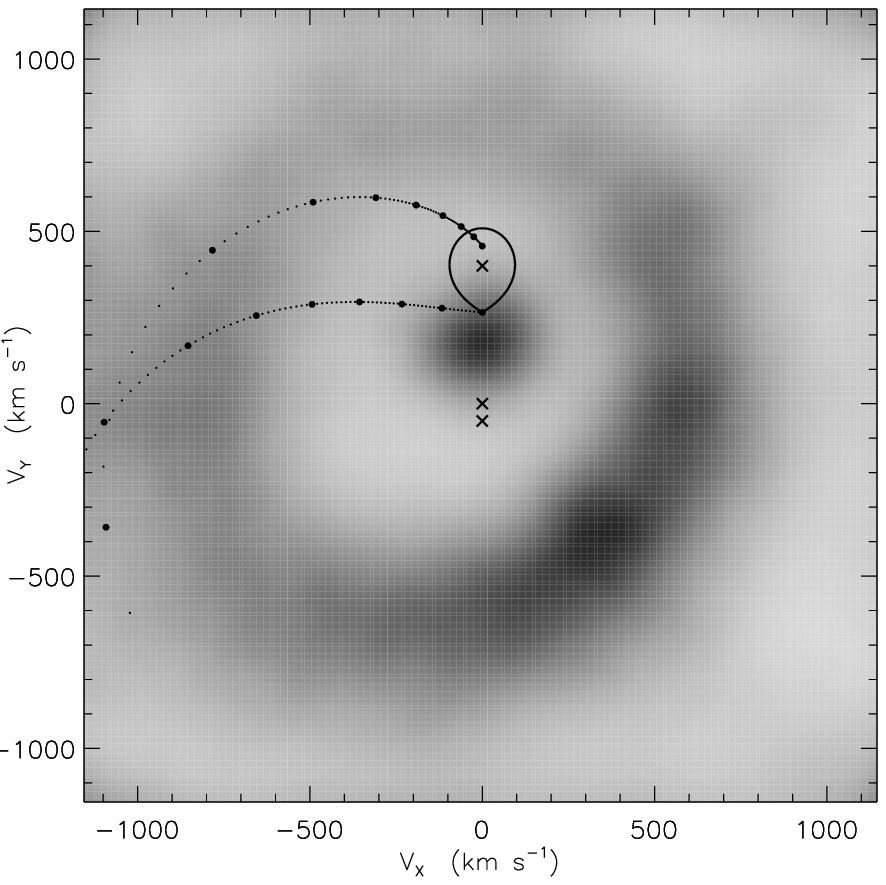}
\caption{\label{fig2} Doppler map of the H$\alpha$ emission line from SDSS\,J0039.}
\end{minipage}\hspace{2pc}%
\begin{minipage}{18pc}
\includegraphics[width=18pc]{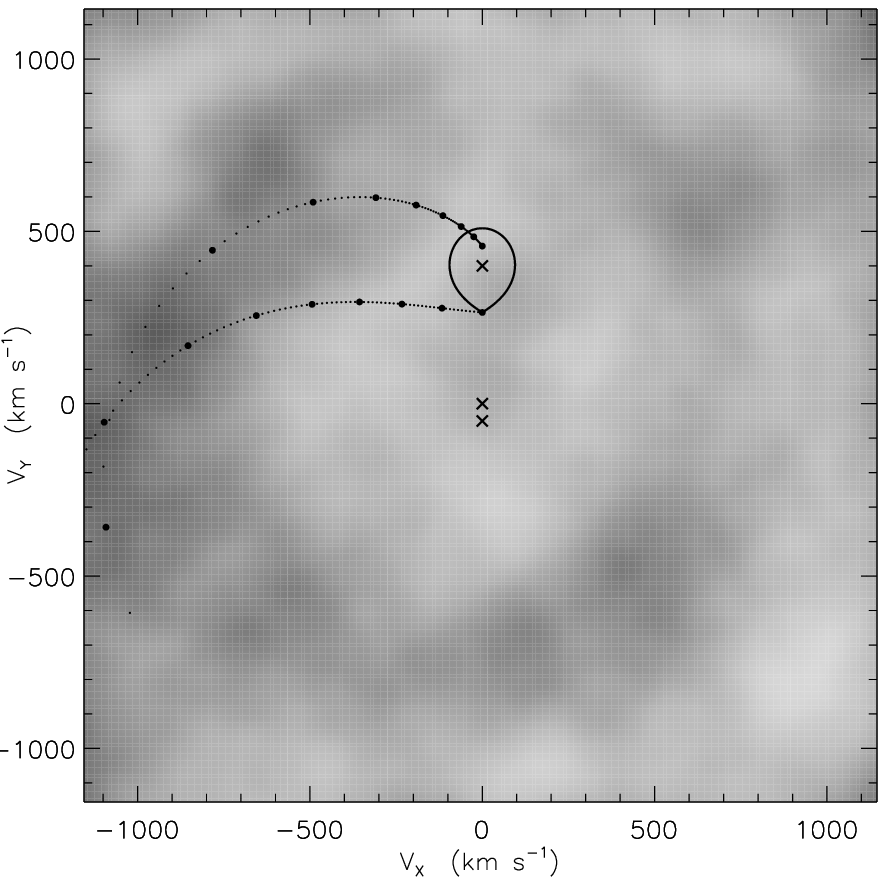}
\caption{\label{fig3} Doppler map of the He\,I 6678\,\AA\ emission line from SDSS\,J0039.}
\end{minipage}
\end{figure}

\begin{figure}[t]
\begin{minipage}{18pc}
\includegraphics[width=18pc]{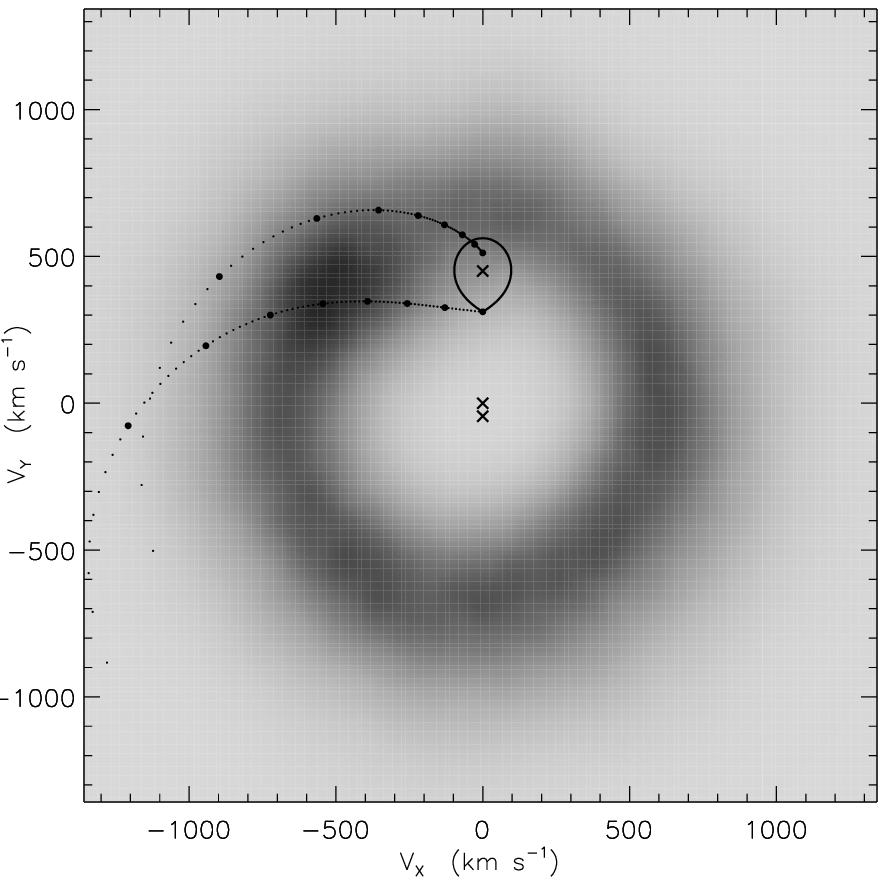}
\caption{\label{f4} Doppler map of the H$\alpha$ emission line from SDSS\,J2205.}
\end{minipage}\hspace{2pc}%
\begin{minipage}{18pc}
\includegraphics[width=18pc]{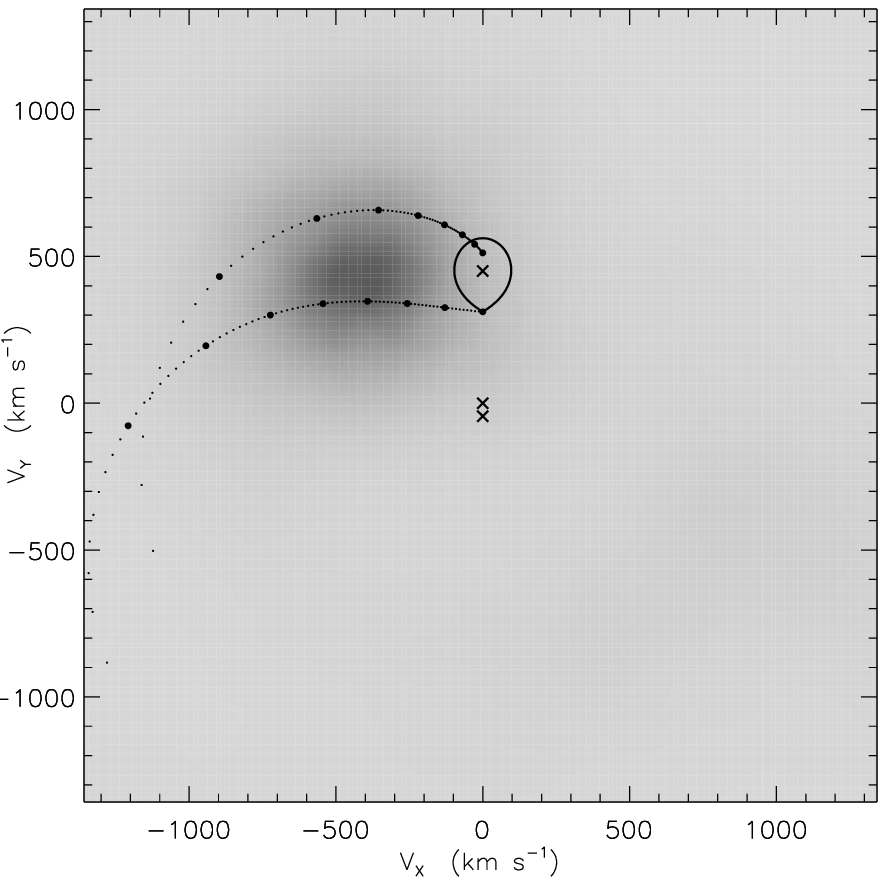}
\caption{\label{f5} Doppler map of the He\,I 6678\,\AA\ emission line from SDSS\,J2205.}
\end{minipage}
\end{figure}

\end{document}